\documentclass[twocolumn,showpacs,amsmath,amssymb]{revtex4}

\input epsf

\def\gsim{\mbox{~{\raisebox{0.4ex}{$>$}}\hspace{-1.1em}
	{\raisebox{-0.6ex}{$\sim$}}~}}

\def\Eq#1{Eq.\ (\ref{#1})}
\def\centerbox#1#2{\centerline{\epsfxsize=#1\textwidth \epsfbox{#2}}}

\def\ca{C_{\rm A}}
\newcommand{\be}{\begin{eqnarray}}
\newcommand{\ee}{\end{eqnarray}}
\def\st{\begin{equation}}
\def\stp{\end{equation}}
\newcommand{\non}{\nonumber\\}
 \def\p{{\bf p}}
 \def\k{{\bf k}}
 \def\x{{\bf x}}
 \def\u{{\bf u}}
 \def\h{{\bf h}}
 \def\q{{\bf q}}
 \def\F{{\bf F}}
 \def\ansatz{{\it Ansatz}}
 \def\mD{m_{\rm D}}
 
\def\alphas{\alpha_{\rm s}}
 \newcommand{\GeV}{\hbox{GeV}}
\def\nc{N_{\rm c}}
\def\nf{N_{\rm f}}

\usepackage{graphicx}

\begin{document}
\title{Shear viscosity in weakly coupled ${\cal N} {=} 4$ Super Yang-Mills
 theory compared to QCD}

\author{Simon C.\ Huot, Sangyong Jeon, and  Guy D.\ Moore}

\affiliation{Physics Department, McGill University, 3600
rue University, Montr\'eal, QC H3A 2T8, Canada}%

\begin{abstract}
We compute the shear viscosity of weakly coupled ${\cal N}{=}4$
supersymmetric Yang-Mills (SYM) theory.  Our result for $\eta/s$, the
viscosity to entropy-density ratio, is many times smaller than the
corresponding weak-coupling result in QCD.  This suggests that $\eta/s$
of QCD near the transition point is several times larger than the
viscosity bound, $\eta/s \geq 1/4\pi$.
\end{abstract}


\maketitle

The experimental study of heavy ion collisions at the Relativistic
Heavy Ion Collider has provided remarkable and sometimes surprising
experimental results.
One prominent surprise is the large size of radial and elliptic flow
observed in the collisions \cite{elliptic_experiment}.  Indeed,
the strength of the elliptic flow observed
in non-central heavy ion collisions
is as large as in models where the
plasma in the early stages after the collision is an
ideal fluid \cite{elliptic_theory}.  
(A fluid is nearly ideal if local equilibration proceeds on
length and time scales small compared to the scale of inhomogeneity of
the system.)  This is a surprise because the plasma droplet
created in heavy ion collisions is not much larger than its intrinsic
microscopic length scale--roughly, the de Broglie wavelength of a typical
excitation.
A fluid's nonideality is set by $(l_{\rm intrinsic}/l_{\rm
  inhom.})(\eta/s)$, where $\eta$ is the shear viscosity and $s$ is the
  entropy density.
%
%
($\eta/s$ is dimensionless in units with 
$\hbar=1=k_{_{\rm B}}$).  Therefore, $\eta/s$ for the quark-gluon plasma
must be 
numerically small to display nearly ideal fluid behavior; a value
$\eta/s >0.2$ 
may be enough to reduce the elliptic flow below what is
observed \cite{Teaney} (though the data may actually
require some non-ideality \cite{need_visc}).

In weakly coupled QCD, which should be valid at sufficiently high
energy densities, $\eta/s$ is relatively large,
$\eta/s \sim 1/\alphas^2 \ln \alphas^{-1}$.  However, the temperatures
achieved at RHIC (probably below $0.5\,\GeV$) are such that the
(running) coupling is not small.  Attempts to interpolate
\cite{Kapusta} between the behavior of $\eta/s$ at high temperature
\cite{AMY} and low temperature \cite{pion_eta} suggest $\eta/s \sim
1$ at the relevant temperatures, which may be too high.
Unfortunately, the only first-principles technique we have to
calculate the actual behavior in this strongly coupled region is
lattice QCD, which is fraught with large uncertainties
when extracting real-time behavior such as $\eta$ \cite{Aarts}.

 Recently, a new theoretical perspective arose.
 There is a theory closely related to QCD, namely ${\cal N}{=}4$
 supersymmetric Yang-Mills theory (SYM), where calculations can be
 performed in the limit of a large number of colors $\nc$ and strong
 't Hooft coupling $\lambda \equiv g^2 \nc$ by
 using string theory techniques.  (The 't Hooft coupling correctly
 accounts for the effective coupling strength given the large number
 of degrees of freedom involved.)   The shear viscosity of SYM theory
 for large $\nc$ and $\lambda$ has been computed; expressed as the ratio
 $\eta/s$, it is \cite{Policastro}
 \st
 {\eta\over s} = {1\over 4\pi}
 \left(1 + {135\,\zeta(3)\over 8 (2\lambda)^{3/2}} + \cdots\right) \,,
 \label{eq:kss}
 \stp
small enough for the fluid to behave almost ideally.

It has further been conjectured by
Kovtun, Son, and Starinets, after evaluating $\eta/s$ in several related
strongly-coupled theories, that $\eta/s=\frac{1}{4\pi}$ is in fact a
lower bound on $\eta/s$ in all systems \cite{viscosity_bound}.
Together with the formal similarities between SYM and QCD, and the
expectation that QCD is strongly coupled at the temperatures relevant in
heavy ion collisions, this has led to a belief that QCD nearly
saturates this bound, and that strongly coupled SYM may resemble QCD
near the transition \cite{believers} and
may be useful for describing other properties of strongly coupled QCD
\cite{heavy_quark,energy_loss,photons}.
This argument is also supported by thermodynamic information; lattice
calculations of the pressure of QCD \cite{Karsch} show that, at a few
times the transition temperature, the pressure is close to $\frac 34$ of
the value at zero coupling, exactly the ratio obtained in strongly
coupled SYM theory \cite{Klebanov}.

To explore whether QCD saturates the
viscosity bound, we think it is useful to examine more carefully how
much SYM really behaves like QCD.  In particular, there is a regime
where calculations can be carried out in both theories; weak coupling.
How close are the values of $\eta/s$ in SYM theory and in QCD at weak
coupling?  And is the physics which sets the viscosity in the two
theories the same?  This paper will address this question.

At weak coupling, a gauge theory plasma behaves much like a gas of
quasiparticles.  A rough estimate of the viscosity and entropy density
are
 \st
 \eta 
 \sim l_{\rm mfp}\, \bar{v}\, (P {+} \varepsilon)
 \sim l_{\rm mfp}\, nT \,, \quad
 s \sim n
 \stp
with $n$ the number density of excitations and $T$ the temperature.
Hence, $\eta/s$ is a measure of the ratio of the mean free path for
large angle scattering, $l_{\rm mfp}$, to the thermal length
$1/T$.  In practice $l_{\rm mfp}$ is momentum dependent and scatterings
do not fully randomize a particle's direction.  To turn the estimate
into a calculation, we must use kinetic theory (Boltzmann equations).
The single particle distribution function $f(\p,\x)$ evolves according
to 
 \st
 \left(\partial_t + {\bf v}{\cdot}\partial_{\bf x}
 \right) f^a(\p, \x, t)
 = -{\cal C}_a[f] \, ,
\label{eq:Boltzmann}
 \stp
with ${\cal C}_a[f]$ a collision term we explain below.  Shear viscosity is
relevant when the equilibrium distribution 
$f_0(\p,\x,t)=(\exp[\gamma(E{+}\u\cdot \p)/T]\pm 1)^{-1}$ varies in
space, $\partial_i u_j \neq 0$.
The stress tensor
(in the local fluid frame) will be shifted by an amount proportional to
this velocity gradient and $\eta$ is the proportionality constant,
$T_{ij}-T_{ij,{\rm eq.}} = 
-\eta (\partial_i u_j{+}\partial_j u_i{-}\frac 23 \delta_{ij} \partial_k
u_k) - \zeta \delta_{ij} \partial_k u_k$.  (We will not discuss the bulk
viscosity $\zeta$.)  The
lefthand side of \Eq{eq:Boltzmann}, at lowest order in $\partial_i u_j$,
reads
\st
(\partial_i u_j) f_0(1{\pm} f_0) \frac{p_i p_j}{E T} = -{\cal C}_a[f] \,,
\stp
while the relevant entry in the stress tensor is
\st
T_{ij} - T_{ij,{\rm eq.}} = \!\sum_a \!\int\! \frac{d^3 p}{(2\pi)^3} \frac{p_i
    p_j}{E} \left( f_a(\p,\x) {-} f_0(\p,\x) \right) \, .
\stp
Both expressions involve the combination $p_i p_j/E$.
The collision term is linear in the departure from
equilibrium and must be inverted to solve for $f-f_0$.

Arnold, Moore, and Yaffe have shown how the inversion of the collision
term can be performed in QCD by variational methods to determine the
viscosity \cite{AMY}.  We follow their treatment, leaving the full
details to the references.  For each species, one must model the
departure from equilibrium by a several parameter \ansatz, with trial
functions $\phi^{(m)}(p)$.  To find the viscosity, one needs the
integral of $\p^2/E$ against each trial function,
$S_{am} \equiv \nu_a\int_p (\p^2/E) \phi^{(m)}(p) f_0(1{\pm}f_0)$ ($\nu_a$ the
multiplicity of species $a$)
and the integral moments of the collision operator,
$C_{am,bn}=\int_p \nu_a \phi^{(m)}(p) 
{\cal C}_a[f_b(k)=f_0+f_0(1{\pm}f_0)\phi^{(n)}(k)
P_2(\cos\theta_{pk})]$, with
$P_2(\cos\theta_{pk})$ the second Legendre polynomial of the angle
between $\p$ and $\k$.  The viscosity is
\cite{AMY}
\st
\eta = \frac{S_{am} C^{-1}_{am,bn} S_{bn}}{15 T}
\stp
where $(am)$ is treated as a single index and $C$ as a matrix.
Their results in massless QCD, divided by the leading-order entropy
density $s=2\pi^2 T^3 g_*/45$ (with $g_*$ the number of bosonic fields
plus 7/8 the number of fermionic fields, $16{+}36(7/8)=47.5$ for
3-flavor QCD), are
 \st
 \frac{\eta}{s} \approx \frac{A}
 {\nc^2 g^4 \ln (B/g\sqrt{\nc})
}\, , \;\;
A,B=\left\{\begin{array}{ll} \!34.8\,,\; 4.67 & \nf=0\,, \\
\!46.1\, , \; 4.17 & \nf=3\,. \\ \end{array} \right.
\label{eq:AMY}
 \stp
Massless QCD with $\nf=3$ is the case closest to the real world, because
the temperatures in heavy ion collisions have $m_u,m_d,m_s < T$ but
$m_c \gg T$.

Now we apply their technique to ${\cal N}{=}4$ SYM theory.  This is a
theory containing gauge fields for an SU($\nc$) symmetry, 4 adjoint
Weyl fermions, and 6 real adjoint scalar
fields, with Lagrangian density given in \cite{SYM_lagrangian}.

As in QCD, two types of collision processes are relevant; elastic
$2\leftrightarrow 2$ processes and inelastic effective $1\leftrightarrow
2$ processes.  The latter are really splitting/joining processes induced
by soft scattering in the plasma.  

The vacuum elastic processes have
remarkably simple squared matrix elements, displayed in
Table~\ref{tbl:M2}.  The integral of these matrix elements over possible
external momenta must be performed by numerical quadratures.
Those elements containing $M_t$ or $M_u$ require
bosonic, and $X_{us}$ and $X_{tu}$ require fermionic, self-energy
corrections, which are the same as in QCD \cite{AMY} but with
$\mD^2=2\lambda T^2$ and $m_{\rm F}^2=\lambda T^2/2$.

\begin{table}[t]
\begin{tabular}{|ll|} \hline
$ SS\rightarrow SS \quad$ & $36 (M_t+M_s+M_u)$ \\
$ SG\rightarrow SG$ & $12 M_t$                 \\
$ GG\rightarrow GG$ & $4 (M_t+M_s+M_u)$        \\
$ FS\rightarrow FS$ & $48 M_t + 144 X_{us}$    \\
$ FS\rightarrow FG$ & $48 X_{us}$              \\
$ FG\rightarrow FG$ & $16 M_t + 16 X_{us}$     \\
$ FF\rightarrow FF$ & $64 (M_t+M_s+M_u)$       \\
\hline
\end{tabular}
\caption{\label{tbl:M2}
Matrix elements squared, summed on all external states of given spin,
where $S,F,G$ are spin $0,\frac12,1$ respectively.  Here
$M_t = 2\left[\frac{u^2+s^2}{t^2}+1\right]$,
$X_{us}= 2\left[-\frac{u}{s} -\frac{s}{u} -1\right]$, and
$M_u$ and $M_s$ are $M_t$ with $u\leftrightarrow t$ and
$s\leftrightarrow t$ respectively.  Elements
for $SS\rightarrow GG$, $FF\rightarrow SS$, $FF\rightarrow SG$, and
$FF\rightarrow GG$ can be obtained by crossing.}
\end{table}

The splitting processes are in a way simpler
than in QCD; all particles are in the adjoint representation and all
hard particles have the same dispersion relation:
$E^2 = p^2+\lambda T^2$ for $p^2 \gg \lambda T^2$.  However, there are
more possible splitting processes, because both gauge and Yukawa
interactions can induce splitting.
The total contribution to $C_{am,bn}$ due to splitting
processes is
\st
C_{am,bn} = 
2\sum_{^{ABC}} \nu_{\!_{ABC}} \frac{\lambda^2}{2 (2\pi)^3} \int_0^\infty dp 
                                          \int_{p/2}^p dk \; {\cal I}
\, ,
\stp
\vspace{-0.2in}
\be
{\cal I} &\!\!=\!\!& {\cal J}_{\!_{ABC}}(p,k)\; {\cal F}(p,k)
           \; e^{p/T} f_A(p) f_B(k) f_C(p{-}k) \\
& & {} \times
           [\delta_{aA} \phi^{(m)}_{A}(p) - \delta_{aB}
          \phi^{(m)}_{B}(k) - \delta_{aC} \phi^{(m)}_{C}(p{-}k)]\non
& & {} \times
           [\delta_{bA}\phi^{(n)}_{A}(p) - \delta_{bB}\phi^{(n)}_{B}(k)
	     - \delta_{bC} \phi^{(n)}_{C}(p{-}k)] \nonumber \,.
\ee
Here ${\cal J}$ is the splitting kernel given in Table \ref{tbl:split}
and ${\cal F}$ is the solution to the integral equation
\be
{\cal F}(p,k) &\!\!=\!\!& \lambda^{-1} \int \frac{d^2 \h}{(2\pi)^2} 
        2\h \cdot \F(\h) \\
\!\! 2\h- i\delta E \F(\h) &\!\!=\!\!&
           \int \frac{d^2 \q}{(2\pi)^2} 
           \frac{\lambda^2 T^3}{\q^2 (\q^2{+}2\lambda T^2)}
\times \\ && \qquad \;\; \left[
\sum_{l{=}p,k,p{-}k}  \!\!\!\Big(\F(\h{+}l\q)-\F(\h) \Big)
 \right]
\non
\delta E &\!\!=\!\!& \frac{\h^2}{2pk(p{-}k)}
       +\frac{[p^2{+}k^2{+}(p{-}k)^2]\lambda T^2}{4pk(p{-}k)}
  \,. \nonumber
\ee
The equation for $\F$ accounts for splitting due to multiple scattering,
with $\q$ the transverse momentum exchange due to a single scattering
and $\h$ the non-collinearity between $p$ and $k$, $\h\equiv \p\times
\k$.

\begin{table}
\begin{tabular}{|ccc|} \hline
$ABC$ & $\nu_{_{ABC}}$ & ${\cal J}_{\!_{ABC}}(p,k)$ \\ \hline
$SFF$ & 12 & $p^2 k (p{-}k)/p^3 k^3 (p{-}k)^3$ \\
$FSF$ & 12 & $p k^2 (p{-}k)/p^3 k^3 (p{-}k)^3$ \\
$FFS$ & 12 & $p k (p{-}k)^2/p^3 k^3 (p{-}k)^3$ \\
$GSS$ &  3 & $2k^2(p{-}k)^2/p^3 k^3 (p{-}k)^3$ \\
$SGS$ &  3 & $2p^2(p{-}k)^2/p^3 k^3 (p{-}k)^3$ \\
$SSG$ &  3 & $2p^2 k^2     /p^3 k^3 (p{-}k)^3$ \\
$GFF$ &  4 & $k(p{-}k)(k^2+(p{-}k)^2)/p^3 k^3 (p{-}k)^3$ \\
$FGF$ &  4 & $p(p{-}k)(p^2+(p{-}k)^2)/p^3 k^3 (p{-}k)^3$ \\
$FFG$ &  4 & $pk(p^2+k^2)            /p^3 k^3 (p{-}k)^3$ \\
$GGG$ &  1 & $(p^4{+}k^4{+}(p{-}k)^4)/p^3 k^3 (p{-}k)^3$ \\ \hline
\end{tabular}
\caption{\label{tbl:split}
Splitting kernels for allowed 3-body processes in ${\cal N}{=}4$ SYM
theory.}
\end{table}

\begin{figure}
\centerbox{0.38}{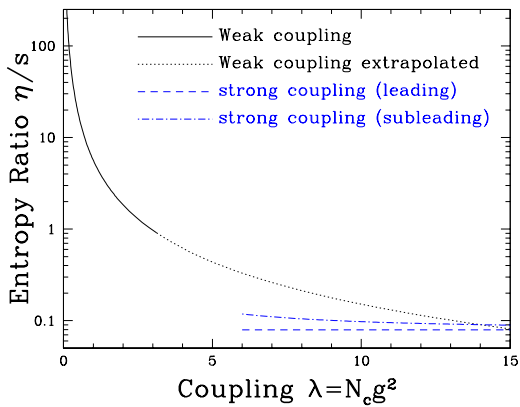}
\caption{ Shear viscosity to entropy density ratio $\eta/s$ in ${\cal
  N}{=}4$ supersymmetric Yang-Mills theory (SYM).  The dotted curve is
  the weak coupling calculation pushed beyond its likely range of
  validity.
}
\label{fig_weakstrong}
\end{figure}

Using the procedure of Ref.~\cite{AMY} and this collision term, we find
that the viscosity at next-to-leading log order is
\st
\frac{\eta_{_{\rm SYM}}}{s_{_{\rm SYM}}} \simeq
\frac{6.174}{\lambda^2 \ln(2.36/\sqrt{\lambda})} \,,
\label{eq:leading_log}
\stp
where as before, $\lambda\equiv \nc g^2$ is the 't Hooft coupling.
At leading order, $\eta/s$ is a complicated function of $\lambda$ which
must be determined numerically.  We resolve the
$O(\sqrt{\lambda})$ ambiguities in its determination
using the procedure of Ref.~\cite{AMY}.  Our ($\nc$ independent!) 
result is plotted in Fig.\
\ref{fig_weakstrong}, which also shows the strong-coupling asymptotic.
The dotted part of the weak-coupling curve is where we believe that
corrections to the weak-coupling calculation may exceed the factor-of-2
level, so the curve guides the eye rather than being a firm
calculation.  (In the one theory where we have an
all-orders calculation of $\eta/s$, namely large $\nf$ QCD
\cite{largeNF}, the leading-order and exact results deviate by about a
factor of 2 when the Debye screening mass $\mD$ reaches the same value
as where we switch to a dotted line in Fig.~\ref{fig_weakstrong}.)
Similarly, the large-coupling asymptotic cannot be trusted
where it is not close to the large $\lambda$ value of $1/4\pi$.  The
curves suggest that strong coupling behavior sets in around
$\lambda \gsim 10$.  Note for comparison that the weak-coupling
expansion for the pressure of SYM theory \cite{SYM_pressure} suggests
that it approaches the strongly-coupled value at a much smaller value of
the 't Hooft coupling, $\lambda \sim 2$.

\begin{figure}
\centerbox{0.40}{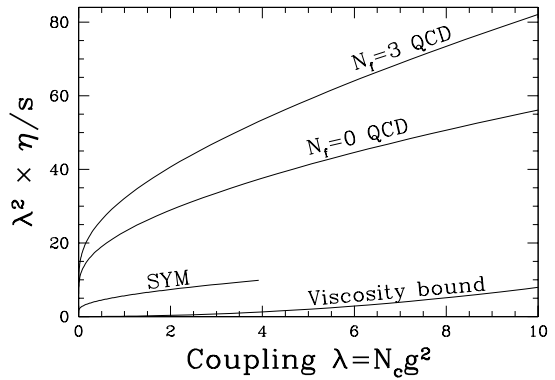}
\caption{$\eta/s$ for SYM theory and for QCD, scaled by the
dominant $\lambda$ dependence and plotted as a function of $\lambda$.
The value in SYM is dramatically smaller than in QCD.}
\label{fig_QCD_SYM1}
\end{figure}

Our result for weakly coupled SYM theory appears rather dramatically
smaller than the result for QCD, both with and without fermions, at
the same coupling, as shown very clearly in Fig.~\ref{fig_QCD_SYM1}.
Naively, this suggests that the viscosity of QCD at
strong coupling should be of order 7 times larger than that of SYM
theory, far from the viscosity bound and closer to the values for other
fluids near critical points.  However, we should explore this conclusion
a little more carefully, to try to understand how this large difference
arose.

The main physics determining the shear viscosity at weak
coupling is Coulomb scattering.
Neglecting all scattering processes but Coulomb scattering
changes the leading-log coefficient $A$ of \Eq{eq:AMY} by less than
$3\%$ ($0.2\%$) for $\nf=3$ QCD (SYM theory).  Working beyond
logarithmic order, neglecting all processes but Coulomb scattering
shifts our viscosity result by $O(25\%)$.  Therefore,
to good approximation the physics we must compare between theories is
the physics of Coulomb scattering.

Two coupling strengths are relevant in Coulomb scattering; the
coupling of a quasiparticle to gauge bosons, and the coupling of
that gauge boson to all other degrees of freedom in
the plasma.  The first coupling (summed over available gauge bosons)
goes as $C_{\rm R} g^2$ with $C_{\rm R}$ the relevant group Casimir.  In
the case of SYM theory, $C_{\rm R}{=}C_{\rm A}{=}\nc$; for QCD it is $\nc{=}3$
for gluons and $(\nc^2{-}1)/2\nc=\frac 43$ for quarks.
The second factor depends on the number, representation, and statistics
of  the other degrees of freedom in the plasma, in exactly the
combination which enters in the Debye screening mass squared.  Therefore
it is natural to expect $s/\eta \sim C_{\rm R} g^2 (\mD^2/T^2)$.

\begin{figure}
\centerbox{0.40}{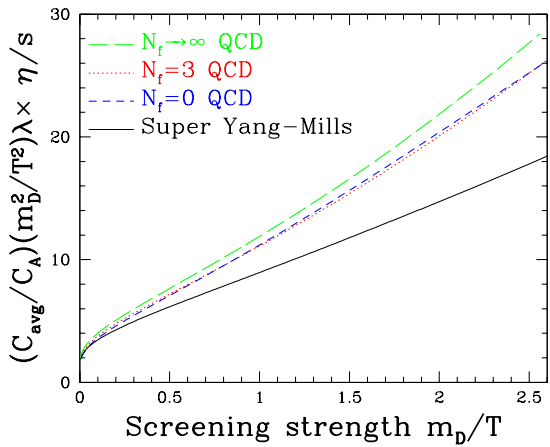}
\caption{Leading-order $\eta/s$ for SYM and for QCD, scaled by 
$C_{\rm avg}\lambda \mD^2/T^2$,
which we argue in the text captures the dominant coupling dependence, as
a function of $\mD/T$.}
\label{fig_QCD_SYM3}
\end{figure}

The quarks in SYM theory are adjoint rather than fundamental, leading to
about a factor of 2 in the Casimir and $\frac 12$ in $\eta/s$.  But much
more importantly, the degree-of-freedom count which enters in $\mD^2$ is
substantially larger in SYM than in QCD.  For instance, for $\nc=3$, SYM
theory has $4\times 2\times 8=64$ fermionic degrees of freedom (four
Weyl fermion species, consisting of a particle and antiparticle in 8
colors)  and $(6{+}2) \times 8 = 64$ bosonic degrees of freedom (6
scalars and 2 gauge boson polarization states times 8 colors), for a
total of 128; while in $\nf=3$ QCD, there are only $3\times 4\times
3=36$ fermionic and 16 bosonic degrees of freedom.  The Debye masses are
correspondingly very different; $\mD^2=2\lambda T^2$ in SYM theory, but
$\mD^2 = \frac 12 \lambda T^2$ in $\nc{=}3$, $\nf{=}3$ QCD.

Since $\eta\propto l_{\rm mfp}$ which scales as $C_{\rm R}^{-1}$ for
each species, a reasonable way to account for the difference in group
Casimirs in a theory like $\nf{=}3$ QCD, with particles in multiple
representations, is to take an average of inverse Casimirs.  Therefore,
define
\st
C^{-1}_{\rm avg} \equiv \frac{C^{-1}_{\rm matter} g_{*{\rm matter}}
+ \ca^{-1} g_{*{\rm adj}}}{g_*} \, ,
\stp
with $C_{\rm matter}$ the Casimir for the representation of matter
fields and $g_{*{\rm matter}}$, $g_{*{\rm adj}}$ the contributions of
each type of field to $g_*$ defined earlier.
Scaling $\eta$ by $C_{\rm avg} \mD^2/\ca T^2$ greatly improves the
agreement between different theories, as shown in Fig.\
\ref{fig_QCD_SYM3}.  However, $\eta/s$ in SYM remains below that in QCD,
because SYM theory has interactions (Yukawa and scalar 4-point) which
are absent in QCD and which introduce additional scattering channels,
further lowering $\eta$.  While these do not contribute at leading-log,
they become more important as the coupling increases.

To conclude, weak-coupling comparisons of QCD with ${\cal N}{=}4$
super Yang-Mills theory strongly suggests that QCD will not
approach the viscosity bound, $\eta/s \geq 1/4\pi$, close to the QCD phase
transition or crossover point.  At weak coupling, SYM theory has
a value of $\eta/s$ which is about 1/7 that of QCD.  This difference
arises because quarks have a smaller coupling to gluons than the gluon
self-coupling, since they are in a different group representation; and
more significantly, because SYM theory has many more degrees of freedom
available as scattering targets than does QCD.

Finally, we comment that both the size of $\eta/s$ and the reliability
of thermal perturbation theory seem to rely on $\mD^2/T^2$ rather than
on $\lambda$ directly; the combination which includes a count of the
degrees of freedom in the plasma is more relevant to thermal behavior.
Relating couplings in this way, $\alphas=0.5$ corresponds to
$\lambda=4.7$, where SYM has thermodynamics very close to the
strongly-coupled behavior but $\eta/s$ still nearly 10 times larger.  It
might be reasonable to believe that QCD plasmas relevant at RHIC fall in
this region.

\centerline{\bf Acknowledgements}

This work was supported in part by 
the Natural Sciences and Engineering Research Council of Canada,
and by le Fonds Nature et Technologies du Qu\'ebec.

\end{document}